\documentclass[pdflatex,sn-mathphys-num]{sn-jnl}
\usepackage{graphicx}%
\usepackage{multirow}%
\usepackage{amsmath,amssymb,amsfonts}%
\usepackage{amsthm}%
\usepackage{mathrsfs}%
\usepackage[title]{appendix}%
\usepackage{xcolor}%
\usepackage{textcomp}%
\usepackage{manyfoot}%
\usepackage{booktabs}%
\usepackage{algorithm}%
\usepackage{algorithmicx}%
\usepackage{algpseudocode}%
\usepackage{listings}%
\usepackage[utf8]{inputenc}%
\usepackage{amssymb}
\theoremstyle{thmstyleone}%
\def\BibTeX{{\rm B\kern-.05em{\sc i\kern-.025em b}\kern-.08em
    T\kern-.1667em\lower.7ex\hbox{E}\kern-.125emX}}

\theoremstyle{thmstyletwo}%

\theoremstyle{thmstylethree}%

\begin{document}

\title[Article Title]{Predicting Stellar Metallicity: A Comparative Analysis of Regression Models for Solar Twin Stars}

\author[1]{\fnm{Sathwik} \sur{Narkedimilli}}\email{21bcs103@iiitdwd.ac.in}
\author[2]{\fnm{Satvik} \sur{Raghav}}\email{satvikraghav007@gmail.com}
\author[1]{\fnm{Sujith} \sur{Makam}}\email{21bcs061@iiitdwd.ac.in}
\author[2]{\fnm{Prasanth} \sur{Ayitapu}}\email{payitapu@gmail.com}
\author*[3]{\fnm{Dr. Aswath} \sur{Babu H}}\email{aswath@iiitdwd.ac.in}

\affil[1]{\orgdiv{Department of Computer Science}, \orgname{Indian Institute of Information Technology (IIIT) Dharwad}, \orgaddress{\city{Dharwad}, \country{India}}}
\affil[2]{\orgdiv{Department of Electronics and Communication Engineering}, \orgname{Amrita School of Engineering}, \orgaddress{\city{Bengaluru}, \country{India}}}
\affil*[3]{\orgdiv{Department of Arts, Science, and Design}, \orgname{Indian Institute of Information Technology (IIIT) Dharwad}, \orgaddress{\city{Dharwad}, \country{India}}}

\abstract{The research focuses on determining the metallicity ([Fe/H]) predicted in the solar twin stars by using various regression modeling techniques which are, Random Forest, Linear Regression, Decision Tree, Support Vector, and Gradient Boosting. The data set that is taken into account here includes Stellar parameters and chemical abundances derived from a high-accuracy abundance catalog of solar twins from the GALAH survey. To overcome the missing values, intensive preprocessing techniques involving, imputation are done. Each model will subjected to training using different critical observables, which include, Mean Squared Error(MSE), Mean Absolute Error(MAE), Root Mean Squared Error(RMSE), and R-squared(R²). Modeling is done by using, different feature sets like temperature: effective temperature(Teff), surface gravity: log g of 14-chemical- abundances namely, (([Na/Fe], [Mg/Fe], [Al/Fe], [Si/Fe], [Ca/Fe], [Sc/Fe], [Ti/Fe], [Cr/Fe], [Mn/Fe], [Ni/Fe], [Cu/Fe], [Zn/Fe], [Y/Fe], [Ba/Fe])). The target variable considered is the metallicity ([Fe/H]).

The findings indicate that the Random Forest model achieved the highest accuracy, with an MSE of 0.001628 and an R-squared value of 0.9266. The results highlight the efficacy of ensemble methods in handling complex datasets with high dimensionality. Additionally, this study underscores the importance of selecting appropriate regression models for astronomical data analysis, providing a foundation for future research in predicting stellar properties with machine learning techniques.}

\keywords{Stellar Metalicity, Random Forest, Linear Regression, Decision Tree, Support Vector, and Gradient Boosting}

\maketitle

\section{Introduction}\label{sec1}

Stellar metallicity is an important, commonly measured parameter, usually represented by a number, of so-called [Fe/H]. It is a basic element of astrophysics that governs the way stars and galaxies have formed and it is important for understanding their histories. Stellar metallicity can serve as a predictor of the age of stars. This paper focuses on predicting the metallicity of solar twin stars (this is a collection of stars that have many important characteristics resembling the Sun) through the application of regression models.

This study utilizes a dataset from the GALAH survey, specifically the high-precision abundance catalog of solar twins, as detailed by Walsen et al. (2024). The dataset, produced using The Cannon algorithm, comprises approximately 38,320 solar twin stars, providing detailed measurements of stellar parameters such as effective temperature, surface gravity, and metallicity, along with 14 different chemical abundances. The rich and extensive nature of this dataset makes it ideal for training and testing regression models aimed at predicting stellar metallicity.

The goal of this research is to see which of the five regression models (Random Forest, Linear Regression, Decision Tree, Support Vector, and Gradient Boosting) gives a better result in predicting stellar metallicity. An overall approach is used for implementing this project in which preprocessing on the data is handled, including dealing with missing values using imputation. The training of each of the regression models is done next and analyzing results for each reporting metric (Mean Squared Error (MSE), Mean Absolute Error (MAE), Root Mean Squared Error (RMSE), and R-squared (R2)) is performed.

By systematically comparing these models, this study seeks to identify the most effective approach for predicting stellar metallicity. This comparative analysis not only contributes to the field of astrophysics by enhancing the accuracy of stellar parameter predictions but also demonstrates the potential of machine learning techniques in astronomical research. The findings of this research will provide valuable insights into the chemical composition and evolution of solar twin stars, offering a benchmark for future studies in the prediction of stellar properties using regression models.

\section{Literature Review}

F Z Zeraatgari et al. \cite{unknowngaeraerherh} show that we can considerably improve machine learning in predicting metallicity and other properties of galaxies by the use of CatBoost and Wide and Deep Neural Networks (WDNN), using data from the Sloan Digital Sky Survey (SDSS) and AllWISE catalogs. In their 2021 paper, the authors concluded that the use of machine learning to recover astronomical properties from observational quantities has the potential to overcome systematic errors and provide more accurate predictions due to observational uncertainties and errors. The impressive results obtained with little effort and time prove the efficacy of modern modeling methods in dealing with complex astrophysical data.
The results achieved are as follows, using CatBoost, an RMSE for SFR, SM and metallicity are 0.349, 0.209 and 0.109 respectively. For the WDNN, the corresponding results were 0.351, 0.210 and 0.110 respectively

According to Limongi et al., \cite{limongi2024evolution}, the evolution of stars like our Sun, which have a metallicity of one (solar), could be ‘best represented’ by following stars from the same narrow (mid-) mass range, considering only the climax points of their evolution and their final states, such as the supernova progenitor scenarios. The paper then focuses on giving a detailed description of the stages of the physical and chemical changes in these stars, emphasizing the importance of understanding the full history of stellar evolution to be able to correctly predict the final state of a star.

According to Amard et al. \cite{amard2020impact}, sun-like stars’ metallicity plays a pivotal role in their rotational and magnetic evolution. The researchers showed that changing the metallicity of such stars can ‘stir dark waters’ and alter their lifecycle by tuning their magnetic activity and evolution of their spins. The researchers summarise the effects of varying metallicity in stellar dynamics throughout the lifetime of a star.
It was found that stars with less metal rotate faster than their solar metallicity counterparts of the same mass and age but exhibit lower stellar activity levels due to higher Rossby numbers. For example, a star with [Fe/H] = +0.3 has a rotation period of about 27-31 days and a magnetic cycle of around 7.5 years, showing higher magnetic activity than the Sun. Additionally, the models predict that metal-rich stars exhibit a higher level of intrinsic variability and enhanced magnetism, which affects the detection of exoplanets and the habitability zones.

Lehmann et al. \cite{lehmann2022survey} explain the method: With a focus on solar analog stars useful for the characterization of extrasolar planets in the HARPS search for exoplanets, the present work takes further steps towards providing accurate metallicity and other stellar parameters on a small scale. Firstly, different from previous applications, this study uses non-iterative differential equivalent width measurements to estimate metallicity, avoid the impact of systematic uncertainties, and increase the measurements' precision. Then, after characterization of the differential RV method, this work compares measurements for eight well-studied solar analogs to other insights into their metallicity, surface gravity, and temperature, thereby verifying the capability of the differential method to provide precise astrophysical stellar parameters.

Making use of the spectral synthesis techniques, Do et al. \cite{do2018super} examined an example of a star with super-solar metallicity which may serve as a typical representative of stellar populations residing in the extreme environment of the Galactic Centre. The peculiar properties and elemental abundance ratios of these typical stars are of great importance for the studies of galactic evolution. In this article, the authors have discussed the methods and findings in detail.

V Ramachandran et al. \cite{ramachandran2019testing} show, that by undertaking a spectroscopic analysis of OB stars in the Wing of the Small Magellanic Cloud (SMC), one can shed light on massive star evolution, star-formation history, and stellar feedback in low-metallicity environments. Utilizing data from stars in the VLT/FLAMES survey, which undertook the largest spectroscopic survey for massive stars in a (dwarf) galaxy to date, the authors examine the physical properties of these stars and try to deduce their evolutionary stage by examining parameters such as the effective temperature, surface gravity and metallicity, improving our constraints on star formation processes, as well as improving our theoretical models of massive star evolution in conditions that are similar to the early Universe. As the authors highlight, these results help us to constrain the models of stellar evolution and star formation in low-metallicity galaxies.
 
\section{Methodology}

\begin{figure}[htbp]
    \centering
    \includegraphics[width=14cm, height=12cm]{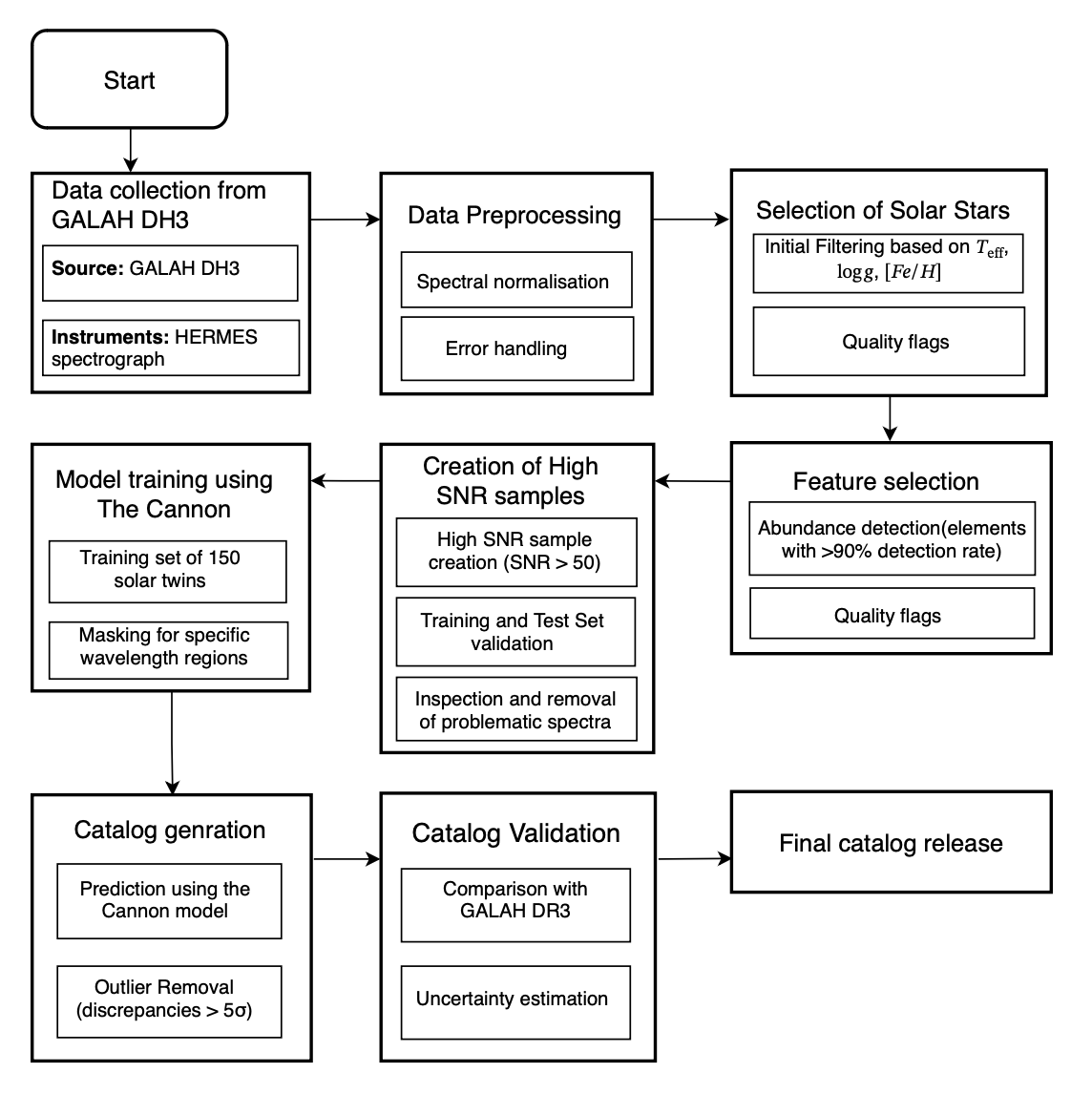}
    \caption{Workflow diagram}
    \label{fig: Architecture of proposed model}
\end{figure}

The workflow illustrated in Figure 1 outlines the comprehensive process of predicting star metallicity from data collection to final prediction. It includes critical stages such as data preprocessing, model selection, and training, evaluation, and comparison, ensuring that the best-performing model is applied for accurate predictions. The structured approach highlights the importance of meticulous preprocessing and rigorous model evaluation in achieving reliable results in astrophysical research.

\subsection{Hardware and Computing tools used}

T4 Graphics Processing Units (GPUs), procured through a cloud-based environment, were used to execute the algorithms, demonstrating high efficiency and reducing computing time.

\subsection{Dataset}

The dataset is carefully cleaned and selected from the bounty catalog (Walsen et al. 2024 \cite{walsen2024assemblinghighprecisionabundancecatalogue}) of solar twins in GALAH as shown in Fig. 2, which contains 38,320 solar twin stars with their stellar parameters and chemical abundances using high-precision measurement of effective temperature (Teff), surface gravity (logg), and metallicity ([Fe/H]), and 14 chemical abundances: [Na/Fe], [Mg/Fe], [Al/Fe], [Si/Fe], [Ca/Fe], [Sc/Fe], [Ti/Fe], [Cr/Fe], [Mn/Fe], [Ni/Fe], [Cu/Fe], [Zn/Fe], [Y/Fe], and [Ba/Fe] together with the internal uncertainty of each parameter. The rich datasets with The Cannon algorithms and with GALAH DR3 solar twins will provide predictive models of stellar metallicity and give life to theories of the chemical enrichment of the Milky Way.

\begin{figure}[htbp]
    \centering
    \includegraphics[width=14cm, height=15cm]{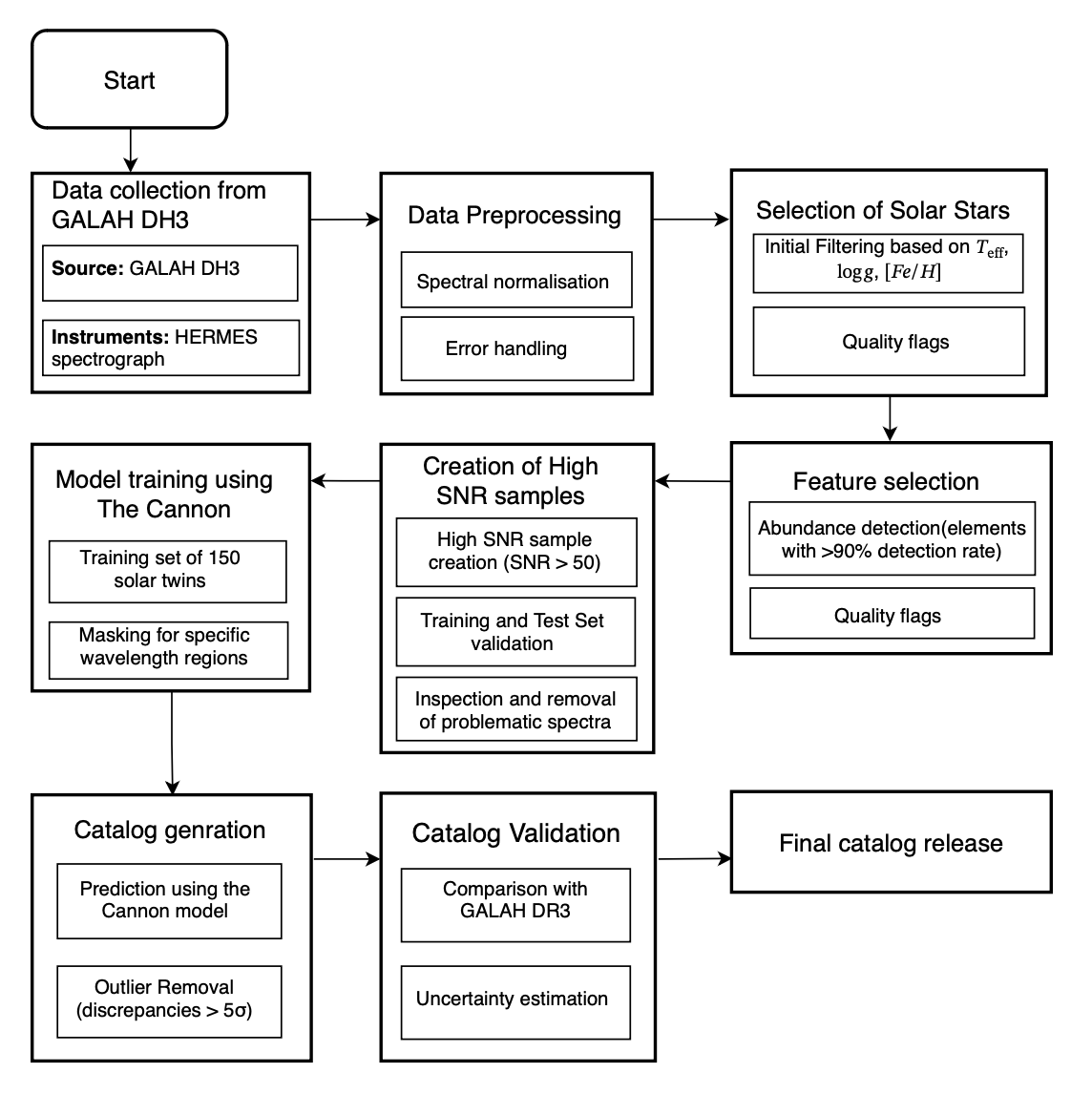}
    \caption{Steps for Solar Twins Catalog Creation}
    \label{fig: Catalog creation steps}
\end{figure}

\subsection{Data Pre-Processing}

Preprocessing is a crucial step in data engineering to ensure the dataset is clean and ready for regression models. We began by removing incorrect entries, duplicates, and users without transactions to avoid noise and bias. Missing numerical values were replaced with the mean of the training set, preserving data integrity. All features were standardized, essential for scale-sensitive models like Support Vector Regression (SVR) and Gradient Boosting. This normalization ensured equal contribution from all features, improved model performance, accelerated training, and reduced the risk of numerical instability, ultimately enhancing the effectiveness and robustness of the regression models.

\subsection{Feature Engineering}

The dataset contained various stellar features such as the effective temperature, surface gravity, and 13 elemental abundances (for instance, Na, Mg, Al, Si, Ca, Sc, Ti, Cr, Mn, Ni, Cu, Zn, Y, Ba) accompanied by the corresponding internal uncertainties of each parameter. All identified correlations between these features and the target metallicity ([Fe/H]) were then used to support feature selection to reduce multicollinearity. Following this, various statistical methods and visualization tools were applied to detect outliers in the dataset and handle them accordingly, such that extreme values would not unreasonably bias the models. Additionally, feature engineering methods have also been applied to create new variables from the original input, especially with exponential polynomials of different order as well as the product or sum of terms to explore nonlinear relationships between variables. Such preprocessing attempts to make the models robust, and the training attempts to fit a model that would best suit the robust dataset, where these attempts can potentially translate into greater reliability in the training models and higher accuracy in the predictions.

\subsection{K-Fold Validation}

We employed 5-fold cross-validation as a rigorous method to evaluate the performance of the proposed regression models. This approach involves dividing the entire dataset into five subsets, or "folds," where the model is trained on four folds and validated on the remaining one. This process is repeated five times, ensuring each fold serves as a validation set exactly once. Performance metrics from each iteration are averaged to provide a single, summarized evaluation metric, offering a robust assessment by reducing variability in the results. The primary advantage of 5-fold cross-validation over a simple train-test split is its ability to provide a more reliable estimate of model performance by averaging results across folds, thereby mitigating variance and bias in the evaluation metrics. Additionally, this method reduces the risk of underfitting and overfitting by exposing the model to different training and validation subsets, facilitating better generalization to unseen data. This approach is particularly beneficial for evaluating regression models across the six combinations of metallicity ([Fe/H]) in solar twin stars.

\subsection{Algorithm Workflow}

 A Random Forest Regressor (RF) was implemented using 100 decision trees to model the relationship between stellar features and metallicity. The dataset, after imputing missing values, was evaluated using 5-fold cross-validation. Performance metrics such as Mean Squared Error (MSE), Mean Absolute Error (MAE), and R-squared (R²) were computed to assess the model's accuracy. The ensemble nature of the RF allowed for the capture of complex non-linear interactions between input features, offering robust performance with a focus on minimising errors and maximising predictive accuracy.

 A Linear Regression (LR) model was applied to the same dataset for baseline comparison, assuming a linear relationship between the stellar features and metallicity. The model was trained using 5-fold cross-validation, and the error metrics (MSE, MAE, RMSE, and R²) were calculated to gauge the model’s performance. Despite its simplicity, LR provided a straightforward interpretation of feature contributions, although its assumption of linearity limited its ability to capture complex interactions in the data.

 A Decision Tree Regressor (DTR) was also trained to predict stellar metallicity, using 5-fold cross-validation for error evaluation. The decision tree model was sensitive to overfitting, yet its ability to model non-linear relationships without requiring feature scaling made it a valuable tool for understanding important features in the dataset. Metrics such as MSE, MAE, and R² were analyzed to evaluate the model's predictive accuracy, with the tree structure providing insights into feature splits.

 The Support Vector Regressor (SVR) was employed to model the data using a radial basis function (RBF) kernel. The SVR model aimed to find a hyperplane in the high-dimensional space that minimizes the error margin. Cross-validation was performed, and error metrics were recorded to assess the model’s capability. Although computationally intensive, SVR was effective in handling non-linear relationships between stellar features and metallicity, especially in cases with limited data points.

 Finally, a Gradient Boosting Regressor (GBR) was implemented to refine predictions through iterative improvements, where each new tree corrected the errors of the previous. The GBR model, evaluated using 5-fold cross-validation, excelled in reducing bias and variance in predictions. Error metrics such as MSE, MAE, RMSE, and R² were recorded, demonstrating the model’s ability to boost performance through gradient optimization, making it particularly effective in modeling complex data patterns in the stellar dataset.

\subsection{Model Selection and Saving the Best Model}

The process involved model selection (choosing between 5 different regression models) to find out which optimal model can predict the metallicity [Fe/H] of solar twin stars the best. The models involved are Random Forest, Linear Regression, Decision Tree, Support Vector, and Gradient Boosting. Each model is designed with a training and evaluation scheme and they are trained using 5-fold cross-validation. At the same time, 5-fold cross-validation is performed for evaluation where performance metrics are averaged over five times and it is a well-known best method that a model can be evaluated thoroughly with this scheme. Meanwhile, Mean Squared Error (MSE), Mean Absolute Error (MAE), Root Mean Squared Error (RMSE), and R-squared (R²) were calculated, and closely evaluated to investigate each model’s predictiveness and accuracy.

\section{Evaluation Metrics}

The evaluations prominently used for the Stellar Mettalicity prediction are MAE, MSE, RMSE, and R2 which are explained as follows.
\subsection{Mean Absolute Error (MAE)}
Mean Absolute Error (MAE) is a metric that measures the average absolute difference between predicted and actual values in a regression model. It provides a straightforward indication of the model's accuracy by averaging the magnitude of the errors, without considering their direction.
\[
\boxed{\text{MAE} = \frac{1}{n} \sum_{i=1}^{n} \left| y_i - \hat{y_i} \right|}
\]

\subsection{Mean Squared Error (MSE)}
Mean Squared Error (MSE) is a metric that measures the average of the squared differences between predicted and actual values in a regression model. It quantifies the variance of the errors, providing a single value that indicates the model's overall accuracy, with larger errors contributing more to the metric.
\[
\boxed{\text{MSE} = \frac{1}{n} \sum_{i=1}^{n} (y_i - \hat{y_i})^2}
\]

\subsection{Root Mean Squared Error (RMSE)}
Root Mean Squared Error (RMSE) is a metric that quantifies the average magnitude of errors between predicted and actual values in a regression model. It is calculated by taking the square root of the average of the squared differences between the predicted and actual values, giving more weight to larger errors.
\[
\boxed{\text{RMSE} = \sqrt{\text{MSE}}}
\]

\subsection{R-squared ($R^2$)}
R-squared (R²) is a statistical measure that indicates the proportion of the variance in the dependent variable explained by the independent variables in a regression model. It ranges from 0 to 1, where a higher value indicates a better fit of the model to the data, meaning the model explains a larger portion of the variability in the observed outcomes.
\[
\boxed{R^2 = 1 - \frac{\sum_{i=1}^{n} (y_i - \hat{y_i})^2}{\sum_{i=1}^{n} (y_i - \bar{y})^2}}
\]

Here, \(y_i\) is the actual value, \(\hat{y}_i\) is the predicted value, and \(n\) is the number of observations.

\begin{equation}
\bar{y} = \frac{1}{n} \sum_{i=1}^{n} y_i
\end{equation}

\section{Implementation and Analysis}
\begin{table}[h]
\caption{Performance Metrics for Various Models}\label{tab:combined_metrics}
\begin{tabular}{@{}lcccc@{}}
\toprule
\textbf{Model} & \textbf{MSE} & \textbf{MAE} & \textbf{RMSE} & \textbf{R²} \\ 
\midrule
Random Forest & 0.0016 & 0.0284 & 0.0403 & 0.9267 \\ 
Linear Regression & 0.0035 & 0.0430 & 0.0587 & 0.8443 \\ 
Decision Tree & 0.0048 & 0.0495 & 0.0689 & 0.7859 \\ 
Support Vector Regression & 0.0219 & 0.1232 & 0.1481 & 0.0120 \\ 
Gradient Boosting Regression & 0.0022 & 0.0348 & 0.0468 & 0.9013 \\ 
\botrule
\end{tabular}
\end{table}

\begin{table}[h]
\caption{Model Accuracy Percentages Corresponding to Metrics}\label{tab:accuracy}
\begin{tabular}{@{}lcccc@{}}
\toprule
\textbf{Model} & \textbf{Accuracy (MSE \%)} & \textbf{Accuracy (MAE \%)} & \textbf{Accuracy (RMSE \%)} \\ 
\midrule
Random Forest & 99.83 & 97.03 & 95.79 \\ 
Linear Regression & 99.64 & 95.51 & 93.87 \\ 
Decision Tree & 99.50 & 94.83 & 92.81 \\ 
Support Vector Regression & 97.71 & 87.14 & 84.54 \\ 
Gradient Boosting Regression & 99.77 & 96.37 & 95.12 \\ 
\botrule
\end{tabular}
\end{table}

Tables 1 and 2 summarise the evaluation metrics (MSE, RMSE, MAE, and R²) and the accuracy percentage by error obtained in the research study for various models.

\subsection{MSE Comparison}
The Mean Squared Error (MSE) is an important metric for rating the performance of the models. It measures the average of the squared difference between the predicted value and the actual value. Random Forest has the lowest value of MSE, as shown in Figure 2 which is 0.0016. This value indicates that Random Forest-Predicted values are closer to the observations (actual values), whereas the predictions done by Support Vector Regression are the worst as is evident from the highest value of MSE as 0.0219 in Table 2. Predictions done with Linear Regression, Decision Tree, and Gradient Boosting Regression are in the middle showing moderate performance. The MSE of Random Forest is very low compared to the other models shown in the graph Fig. 3 which also implies the same fact.

\begin{figure}[htbp]
    \centering
    \includegraphics[width=14cm, height=10cm]{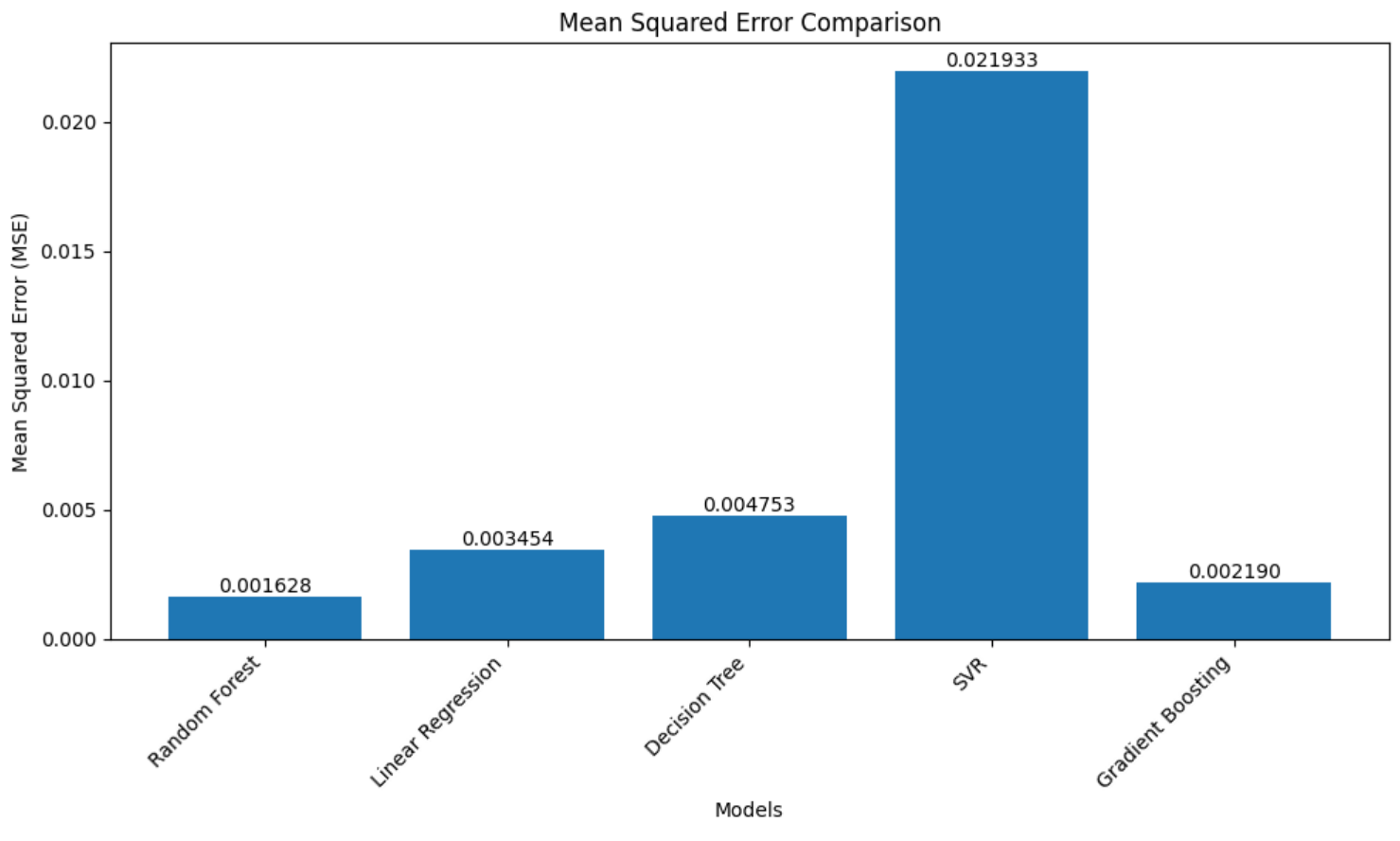}
    \caption{Mean Squared Comparison}
    \label{fig: Mean Squared Comparison}
\end{figure}

\subsection{MAE Comparison}
The Mean Absolute Error (MAE) provides insight into the average absolute differences between predicted and actual values, making it less sensitive to outliers compared to MSE. Random Forest again leads with the lowest MAE of 0.0284, followed closely by Gradient Boosting Regression at 0.0348. Linear Regression and Decision Tree models show higher MAE values, indicating that their predictions deviate more from actual values. SVR, with an MAE of 0.1232, clearly demonstrates the least accuracy, highlighting its struggles in generalization. Fig. 4 illustrates the performance of the models in terms of MAE, emphasizing the superiority of Random Forest.

\begin{figure}[htbp]
    \centering
    \includegraphics[width=14cm, height=10cm]{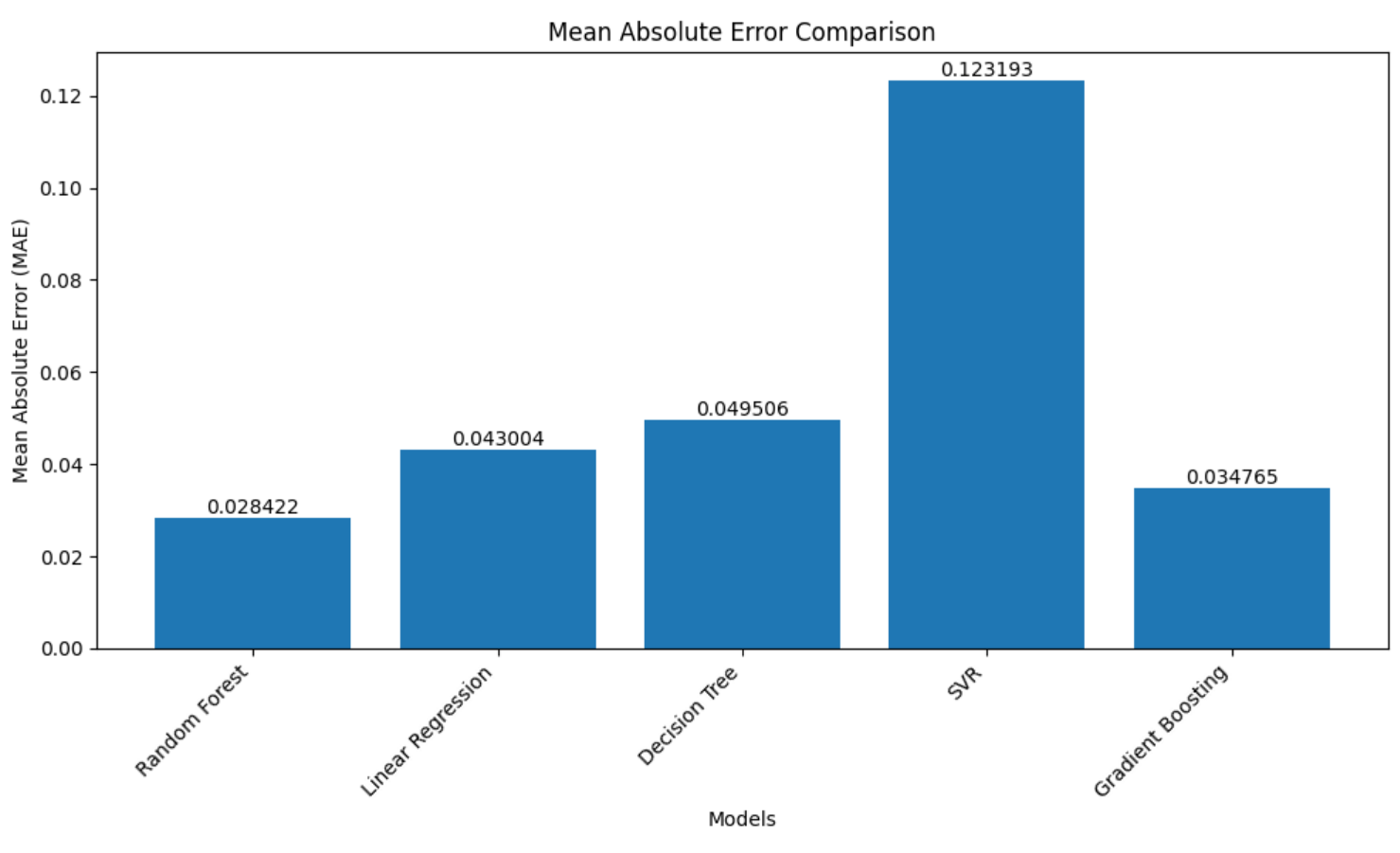}
    \caption{Mean Absolute Error Comparison}
    \label{Mean Absolute Error Comparison}
\end{figure}

\subsection{RMSE Comparison}
The Root Mean Squared Error (RMSE), being the square root of MSE, provides a measure of the standard deviation of the residuals. Random Forest maintains the lowest RMSE at 0.0403, reinforcing its position as the most reliable model among those evaluated. In comparison, SVR shows the highest RMSE of 0.1481, indicating substantial prediction errors. The RMSE values for Linear Regression, Decision Tree, and Gradient Boosting Regression lie in between, reflecting their varying degrees of accuracy. As depicted in Fig. 5, the RMSE metrics further confirm the effectiveness of Random Forest in minimizing prediction errors.

\begin{figure}[htbp]
    \centering
    \includegraphics[width=14cm, height=10cm]{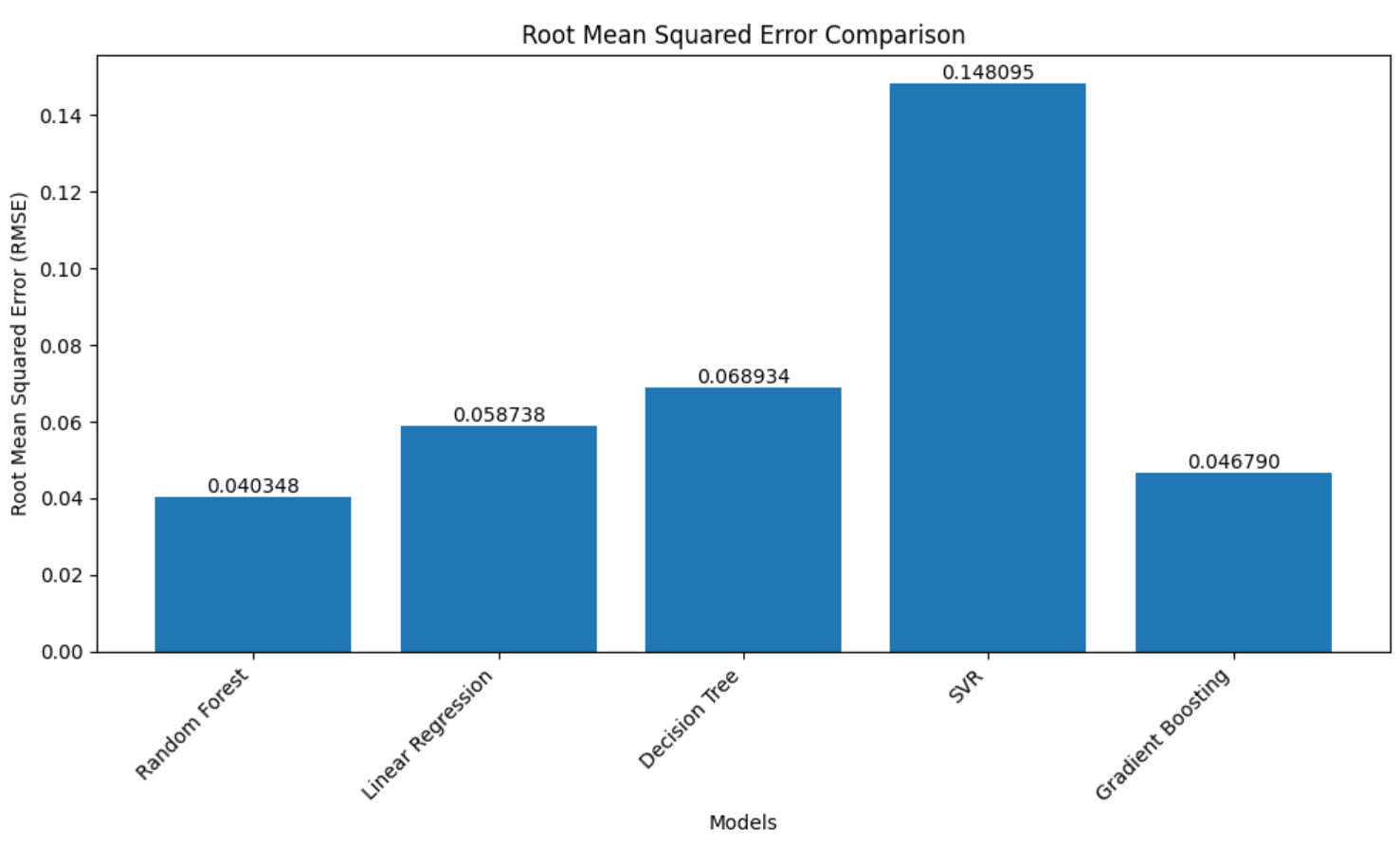}
    \caption{Root Mean Squared Error Comparison}
    \label{Root Mean Squared Error Comparison}
\end{figure}

\subsection{R² Comparison}

\begin{figure}[htbp]
    \centering
    \includegraphics[width=14cm, height=10cm]{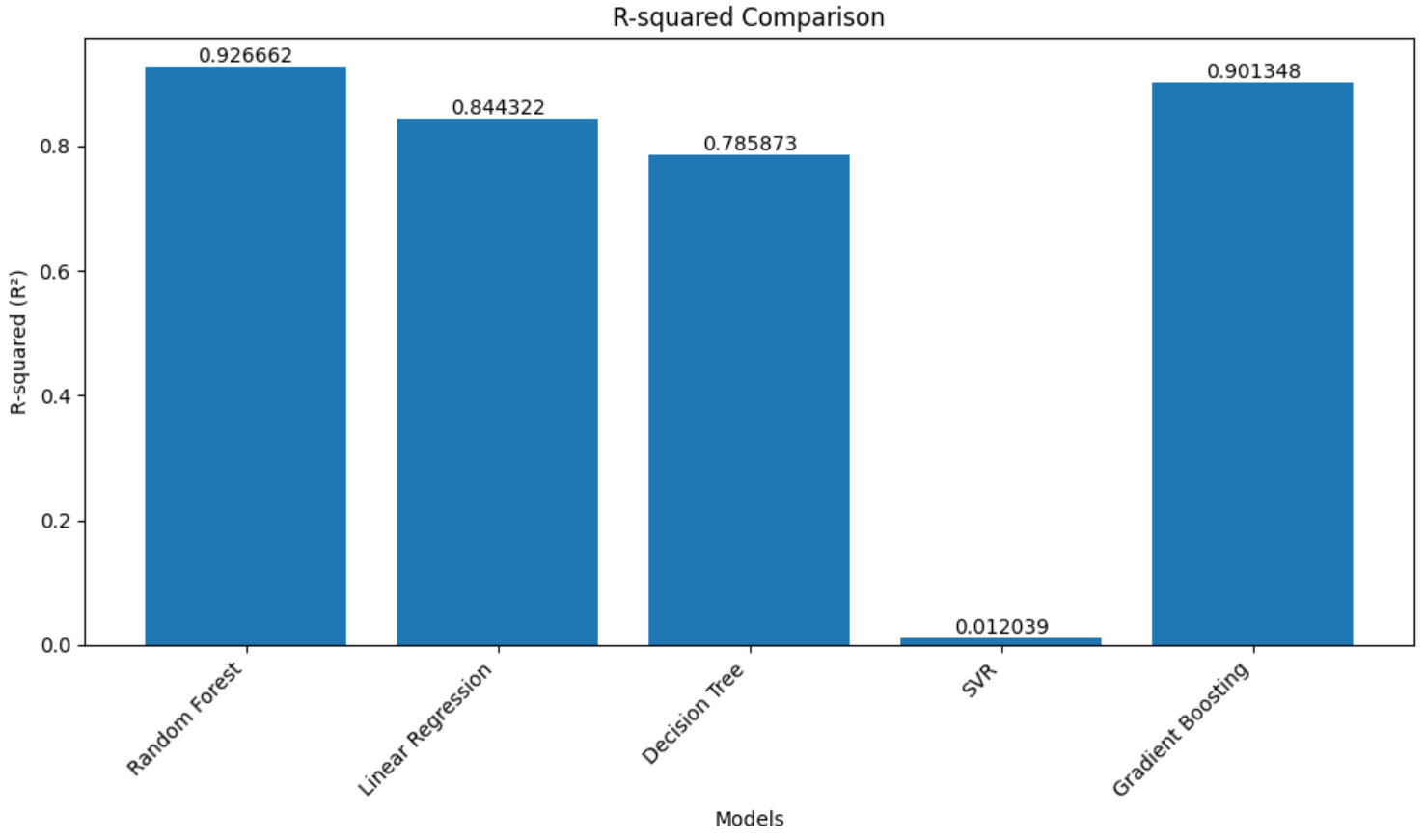}
    \caption{R-squared Comparison}
    \label{R-squared Comparison}
\end{figure}

R-square (R²) is a statistic that represents the proportion of the variance of the dependent variable that is predicted by the independent variables in the model. Random Forest provides an R² score of 0.9267. This is quite an impressive result since an R² of 0.9267 represents 92.67\% of the variance explained. Gradient Boosting Regression takes second place with 0.9013 R², an impressive number, demonstrating a strong performance. Finally, the support vector regression algorithm only manages an R² of 0.0120, an extremely poor result, indicating poor ability to learn the function that best fits the data. Fig 6 below shows the R2 scores for each model.

\subsection{Accuracy Comparison}
Accuracy percentages corresponding to MSE, MAE, and RMSE provide a comprehensive view of model performance. Random Forest leads in all accuracy metrics, with percentages of 99.83\% for MSE, 97.03\% for MAE, and 95.79\% for RMSE. Gradient Boosting Regression also performs well, particularly in MAE and RMSE. In contrast, SVR shows the lowest accuracy across all metrics, indicating its limitations in predictive tasks. Fig. 7 summarizes the accuracy metrics for each model, reinforcing the overall performance trends observed.

\begin{figure}[htbp]
    \centering
    \includegraphics[width=14cm, height=10cm]{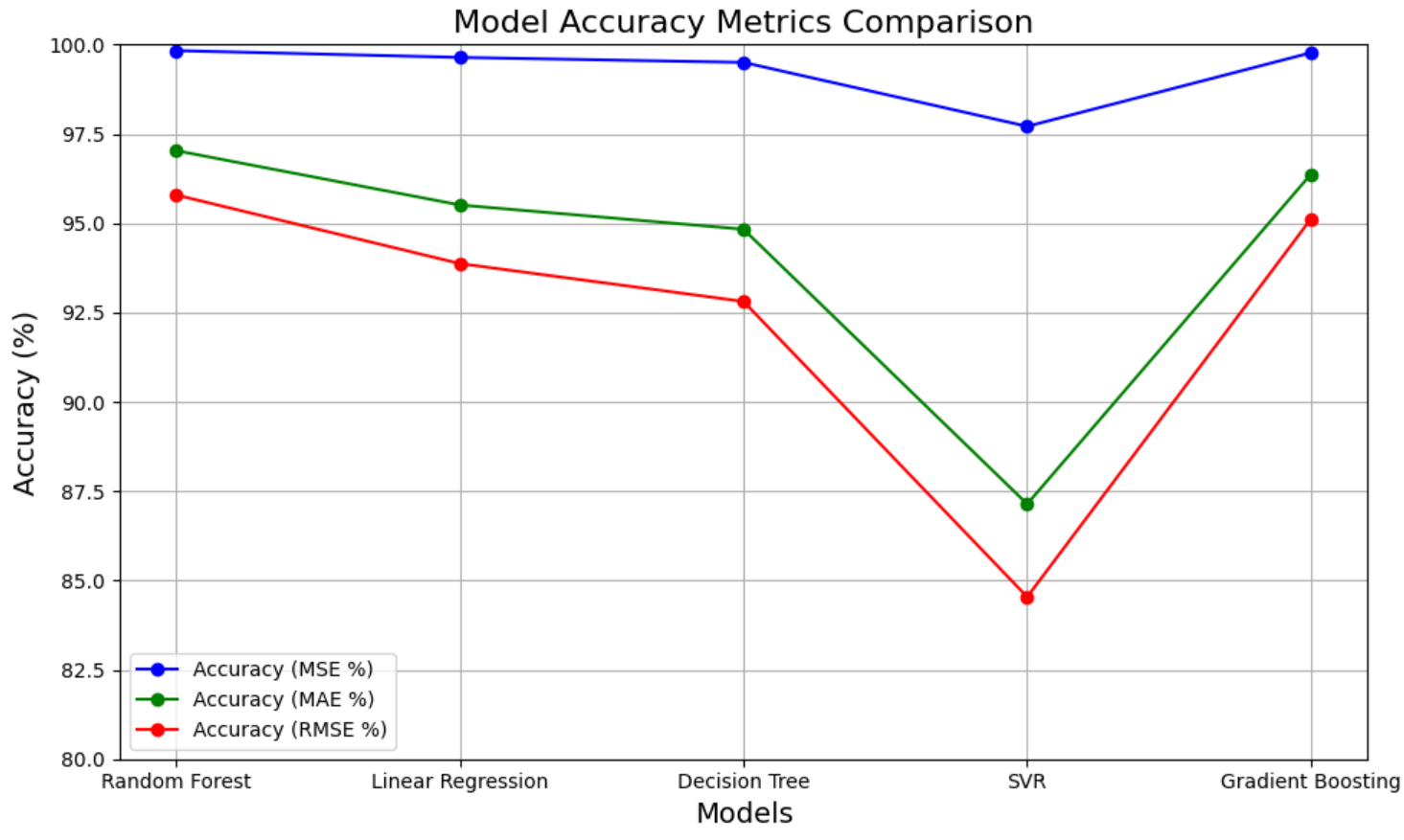}
    \caption{Model Accuracy Metrics Comparison}
    \label{Model Accuracy Metrics Comparison}
\end{figure}

\subsection{Summary}
In summary, the performance metrics reveal distinct strengths and weaknesses across the models evaluated. Random Forest emerges as the top performer, exhibiting the lowest error metrics and a high R² score, indicating its reliability for predictive tasks. Gradient Boosting also shows strong performance, closely following Random Forest. Linear Regression and Decision Tree models demonstrate moderate effectiveness, while Support Vector Regression struggles significantly, suggesting it may not be suitable for this dataset. These insights can inform model selection based on specific performance requirements for future predictive modeling endeavors.

\section{Conclusion}

This study proved that many machine learning regression models can be used to estimate solar twin stars' metallicity [Fe/H]. The best model among these is the Random Forest regressor with the following performance metrics:
Mean Squared Error (MSE) = 0.0016282376367441286
Mean Absolute Error (MAE) = 0.028421949671005153
Root Mean Squared Error (RMSE) = 0.04034774950456317
R-squared (R²) = 0.9266620808330737
The predictive model has a high overall performance in terms of error and accuracy.

In summary, the Random Forest regressor emerged as the most reliable model for predicting stellar metallicity, demonstrating the significance of advanced machine learning techniques in astrophysical research. The study's rigorous preprocessing and imputation methods ensured minimal information loss, thereby enhancing prediction accuracy. These findings underscore the potential of machine learning models in handling high-dimensional astronomical datasets and pave the way for future research to further refine these models and expand their applicability.

\section{Future Scope}

Future studies should incorporate additional photometric data from large-scale surveys to enhance the accuracy and generalizability of machine learning models for star formation rate (SFR) prediction. Utilizing multi-wavelength photometric data, particularly from infrared and ultraviolet bands, can provide crucial insights into various phases of star formation. Furthermore, integrating parameters such as luminosity, radial velocity, and higher-resolution chemical abundances will significantly improve models for predicting stellar metallicity. These additions will help uncover new relationships crucial for understanding star formation and evolution. Advanced machine learning techniques, including ensemble methods like XGBoost and LightGBM, and deep learning architectures like convolutional neural networks (CNNs) and transformers, can further enhance prediction accuracy and model robustness.

Looking forward, temporal analysis of metallicity evolution using time-series analysis and recurrent neural networks (RNNs) may provide deeper insights into stellar evolution. Cross-survey validation with data from stellar surveys like APOGEE or LAMOST can assess the generalizability of predictive models. Additionally, quantum algorithms hold the potential to improve predictive modeling efficiency and accuracy, offering new pathways for astronomical data analysis. Real-time predictive tools for telescopes could enable astronomers to quickly estimate the metallicity of newly observed stars, making scientific decisions faster and more accurate while ensuring robust and reliable estimates of both SFR and stellar metallicity. Integrating these models with astrophysical simulations will create a feedback loop between observation and theory, driving future research in star formation and cosmology.

\bibliographystyle{plain} 

\end{document}